\documentclass[secnumarabic, graphics,floatfix, nofootinbib,tightenlines,nobibnotes, aps, prb, twocolumn]{revtex4-1}

\usepackage[colorlinks=true,citecolor=blue,linkcolor=magenta]{hyperref}
\usepackage{amsmath}
\usepackage{graphicx}
\usepackage{dsfont}
\usepackage{changepage}
\usepackage{endnotes}
\usepackage{fancyhdr}
\usepackage{amsthm, amssymb}
\usepackage{floatflt}
\usepackage{float}
\usepackage{array}
\usepackage[usenames,dvipsnames]{color}
\usepackage{subfig}
\usepackage{chemformula}

\newcommand{\be}{\begin{align}}
\newcommand{\ee}{\end{align}}

\def \be{\begin{equation}}
\def \ee{\end{equation}}
\def \ba{\begin{array}}
\def \ea{\end{array}}
\def \bea{\begin{eqnarray}}
\def \eea{\end{eqnarray}}

\def \>{\rangle}
\def \<{\langle}

\def \e{{\epsilon}}

\def \D{{\Delta}}

\def \w{{\omega}}

\def \e{{\epsilon}}

\def \ba{\begin{align*}}
\def \ea{\end{align*}}

\newcounter{indice}

\begin{document}
\title{Proposal to measure the pair field correlator of a fluctuating pair density wave.}
\author{ Patrick A. Lee }
\affiliation{
Department of Physics, Massachusetts Institute of Technology, Cambridge, MA, USA
}

\date{\today}

\begin{abstract}
I propose a method to directly measure the space and time dependence of the pair field correlator of a pair density wave. The method is based on two separate ideas. First, we adopt the solenoid insertion method of Ref. \onlinecite{kapon2017} to provide  the momentum in a tunnel junction. Second, we suggest the use of optimal or over-doped Bi-2201 films as a tunneling electrode with a known charge ordering wave-vector which can match the expected pair density wave wave-vectors we wish to study. The method is applicable   to both fluctuating and ordered states. Potential applications are to the proposed stripe pair density wave order in LBCO and the possible pair density wave fluctuations at finite temperature or in  high magnetic field in underdoped YBCO as well as
 other members of the cuprate family.
\end{abstract}

\maketitle

\section{introduction.}
\noindent
A pair density wave (PDW) is a superconductor with Cooper pairs which carry a finite momentum $\mathbf{P}$. It is characterized by the order parameter, 
\be
\D_\mathbf{P} = |\D_\mathbf{P}|e^{-i\mathbf{P}\cdot\mathbf{r}+i \phi(\mathbf{r})}
\label{Eq: deltaP}
\ee
In addition we can have a superposition of $\D_\mathbf{-P}$. Furthermore $\mathbf{P}$ can run in several directions to form bi-directional or tri-directional PDW's. The concept was first introduced by Larkin and Ovchinnikov\cite{larkin1965inhomogeneous}  and by Fulde and Ferrell\cite{fulde1964superconductivity} as a way to overcome the Pauli limiting effect of a magnetic field on the superconductor (SC). In recent years PDW has come into prominence in the context of cuprate superconductors, particularly in the underdoped regime. As early as 2002, Himeda, Kato and Ogata \cite{himeda2002stripe} found by projected variational Monte Carlo methods that the PDW is the ground state or very close to the ground state in the presence of stripe order. Starting from the standard stripe picture 
 \cite{tranquada1995jm} of a period 8 spin density wave (SDW) and a period 4 CDW, they found that the d wave superconductor is more stable if the sign of the order parameter is reversed at the hole poor region of the CDW, leading to a period 8 PDW. We shall refer to this state as the stripe-PDW. 
  Strong anisotropy in the transport properties was discovered in the  $\text{La}_{2-x}\text{Ba}_x\text{CuO}_4$ (LBCO) system\cite{li2007two} and explained in terms of stripe-PDW stacked perpendicular to each other to cancel out the interlayer Josephson coupling. \cite{himeda2002stripe,berg2007dynamical}  For a review, see  Ref.\onlinecite{berg22009NTPhysstriped,fradkin2015colloquium}. The current picture is that in 1/8 doped LBCO, 2D superconductivity consisting of stripe PDW appears below about 30K while 3D ordered uniform d wave appears below 5K. It is believed that PDW remains ordered below 5K with perhaps diminished strength. Recent measurement of current phase relation in this low temperature regime confirms this view. \cite{vanharlingen} However, a direct detection of PDW order is still lacking in this material. 

Direct detection of PDW was reported in a local Josephson  tunneling probe measurement using STM on the surface of Bi-2212. \cite{hamidian2016detection} It is known that static charge density wave with wave-vector Q close to 1/4 reciprocal lattice unit (r.l.u.) exists in these materials.  The experiment observed modulation of the pairing order parameter at wave-vector Q, co-existing with the uniform order. This is expected based on Landau theory, because the combination of uniform pairing and charge order will induce PDW with the same period. Nevertheless, this is an experimental tour-de-force and to our knowledge the first direct measurement of PDW order. However, this measurement can detect static order only and not applicable to a fluctuating PDW.

In 2014, the idea of fluctuating PDW was proposed by Lee \cite{lee2014amperean} as the "mother state" behind the pseudo-gap phenomenon in underdoped cuprates. This was based mainly on analysis of a variety of photo-emission, transport, scattering and other data. A somewhat more cautiously phrased proposal that PDW may be the origin of the  energy gap near (0,$\pi$) and the Fermi arc near the nodal direction was suggested earlier based on the stripe PDW picture. \cite{berg22009NTPhysstriped} The general picture is that fluctuating PDW at wave-vector P can induce static or quasi-static charge density wave at Q=2P . (in Bi-2212, the charge density wave period is about 4$a$ ) In this sense the observed charge density waves are subsidiary or composite order. A consequence of this picture is that in the presence of an ordered d wave superconductor, a static PDW will induce charge ordering at wave-vector P=Q/2. Thus the recent  discovery \cite{edkins2018magnetic} of short ranged but static charge order at wave-vector P ( period 8$a$ ) in the vicinity of the vortex core  provided strong support for this point of view. \cite{wang2018,dai2018pair} In at least one of the models,\cite{dai2018pair} the phase gradient near the vortex pins the fluctuating PDW to become static and produce the static induced charge order. Note that this is not a direct measure of the PDW and the measurement also requires the PDW to be static.

Given all the current excitement, it will be great to find a way to directly detect the PDW pair correlator. Since the pairing order is off-diagonal, its fluctuation can only be probed by another off-diagonal system and it was pointed out long ago  by Scalapino 
\cite{scalapino} 
that the correlator $C(r,t)$ = $<\D_P(\mathbf{r}
,t)\D^*_P(0,0)>$ can be probed by the measuring the tunnel conductance as a function of voltage and an applied magnetic field parallel to the junction. It is natural to apply this idea to the PDW and use the parallel magnetic field to provide the momentum to couple to a uniform superconductor probe electrode. \cite{lee2014amperean} However, it was later recognized that if the probe SC is a layered type 2 superconductor like a high $T_c$ cuprate, vortex penetration into the probe SC severely limit the momentum that is accessible. \cite{koren2016}

\begin{figure}[htb]
\begin{center}
\includegraphics[width=3in]{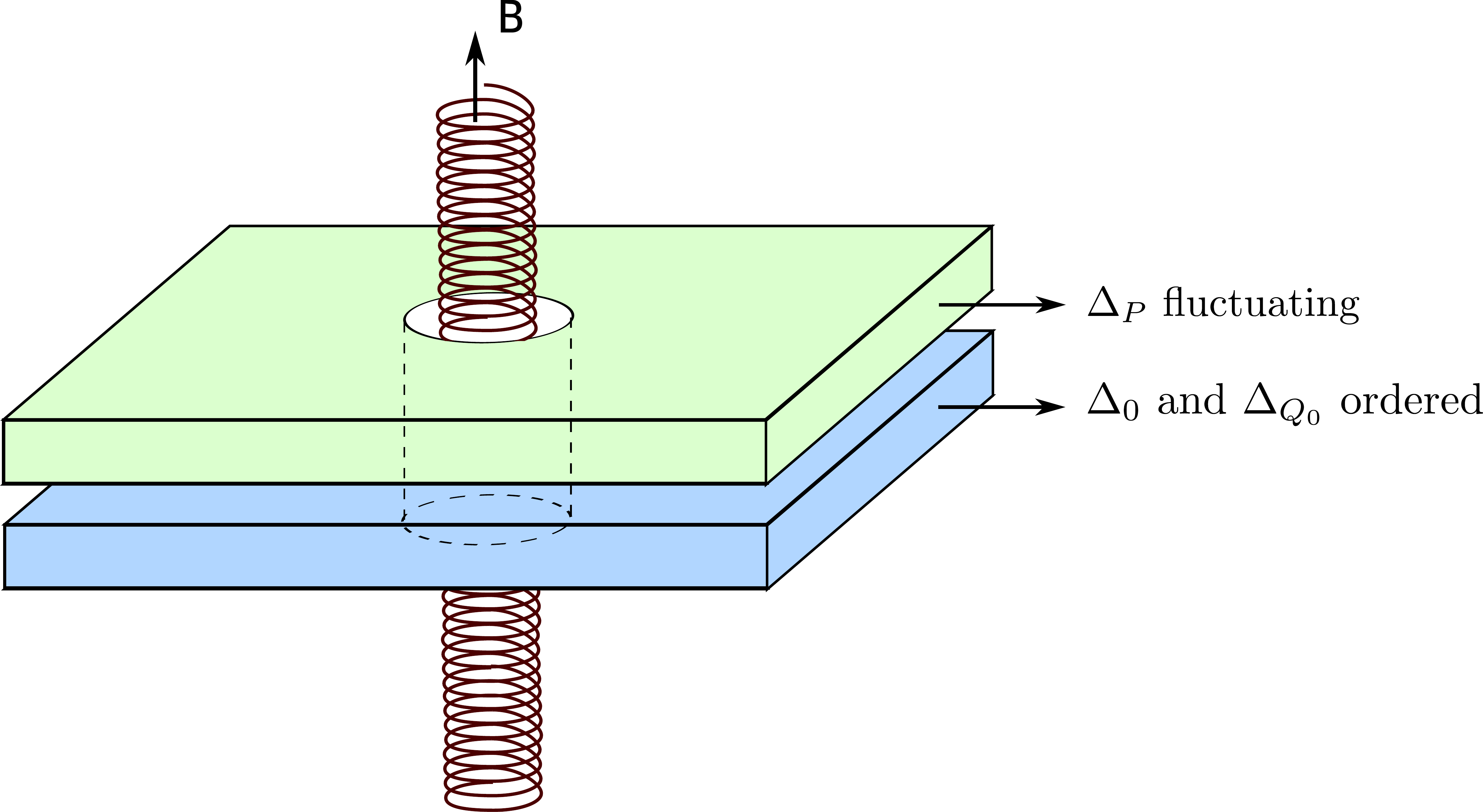}
\caption{A drawing of the proposed scheme to measure PDW fluctuations. The sample being studied forms a tunnel junction with a probe superconductor that has long range ordered uniform or finite momentum pairing, or both. A hole is drilled in the middle and a solenoid  is inserted. (see ref \onlinecite{kapon2017} )}
\label{Fig: setup}
\end{center}
\end{figure}

In this paper we revisit this question in light of two recent experimental developments. First, in a remarkable paper, Kapon et al \cite{kapon2017} developed a new method to measure the stiffness and critical current density by inserted a solenoid into a hole drilled in a superconductor. For a sample much thinner than than hole radius, the magnetic field is mostly confined to the inside of the solenoid and the SC experiences only the vector potential $\bf{A}$ produced by the solenoid. In this way they can create a supercurrent without subjecting the SC to a magnetic field.  As a proof of concept,  a relatively small supercurrent was produced that is sufficient to destroy SC in an LSCO sample 2.1K below its transition temperature of 27.9K. With the use of a superconducting coil, the current through the solenoid  has been 
 increased to 600 mA from 10mA report in ref \onlinecite{kapon2017} without heating and a much larger supercurrent has been achieved, sufficient to drive an LSCO sample normal at 4K. \cite{kerenprivate} Apparently the solenoid current can be increased further.  We propose to extend this method to a tunnel  junction as shown in Fig 1. The idea  to use this geometry instead of the parallel magnetic field to provide the momentum to tunnel into a PDW. While there is no fundamental limit on the size of the vector potential that can be produced, the current must not be so large that the probe SC is destroyed. In section 3 we come to the unfortunate conclusion that even a probe SC with very short coherence length will be killed before  $\bf{A}$ is large enough  to provide the   momentum needed to couple to the PDW in Cuprates.  Based on the idea that the wave-vector is half of the commonly observed charge ordering wave-vector, we expect the PDW wave-vector (in reciprocal lattice unit, r.l.u.) , to be 1/8 for 1/8 doped LBCO and  underdoped Bi-2212 and Bi-2201 and 1/6 for underdoped YBCO. \cite{blanco2014resonant}  The accessible region is sketched in fig 2. 
  
  It is clear to us that in order to access the PDW fluctuation, we need a probe SC that is a PDW with similar wave-vector. Where is this PDW going to come from? Here a second set of experiments comes to the rescue. Recall that a PDW was detected in a Bi-2212
   sample \cite{hamidian2016detection}, so we know that such a PDW exists. However, its wave-vector is 1/4 which is too far from our target wave-vector. On the other hand, it is known that  Bi-2201 is quite unique in that the charge order extend to optimal doping and to the overdoped region. The optimal and overdoped samples have progressively smaller $Q$ vector. \cite{comin2014} 
   Recently a detailed study of the doping dependce of $Q$ was made. \cite{watt} It is found that each sample has a spread of Q vector, but the spread becomes smaller with increasing doping and is centered around 1/6 for optimal doping ($T_c$ = 35K ) and nearly 1/8 for an over-doped ($T_c$ = 15K ) sample. These Q vectors match well the target P for the purported fluctuating PDW. Another way to think about this is simply that the charge order at the interface provides the needed momentum to couple the two SC's. The result is sketched in Fig 3 and we can see that the shifted peak in the conductance now lies within the accessible region.  Furthermore, while the $T_c$ of Bi-2201 is low, the optimal and over-doped samples have very high $H_{c2}$ and should withstand a very large $\bf{A}$. On the other hand, near 1/8 doping the $H_{c2}$ is anomalously small and the superconductivity of the sample is easily destroyed. Thus our
    proposal is to create a tunnel structure using optimally doped or overdoped Bi-2201 as probe electrodes in a setup shown in Fig 1. In practice it may be best to grow such a tunnel structure by MBE, as the layer by layer growth of  Bi-2201 films in combination with other layer materials  has already been demonstrated. \cite{eckstein1995} The proposed experimental protocol is to turn on the solenoid field at a temperature below $T_c$ of the probe SC and measure the tunneling conductance as a function of the solenoid current.  The result is predicted to show a peak shifted by the difference of the wave-vector of the probe SC and the fluctuating PDW. Note that this experiment can be carried out even below $T_c$ of the target sample. While the uniform SC in the probe can couple with the
uniform component in the sample, that Josephson current can easily be killed by a parallel magnetic field, leaving the fluctuation contribution to the conductance. This method can also work in case the PDW in the sample is ordered, as is expected to be the case for LBCO. However, in order to probe the momentum dependence, a special protocol is required, as will be discussed later. 
In section 2 we describe a formulation of the tunneling current in terms of gauge invariant quantities that will facilitate later discussions. Since both electrodes are subject to the same gauge field one might worry that its effect is canceled in the tunneling process. The gauge invariant formulation makes it clear what are the conditions under which the cancellation does not take place. In section 3 will provide details of our analysis of a number of different scenarios.

\begin{figure}[htb]
\begin{center}
\includegraphics[width=3.2in]{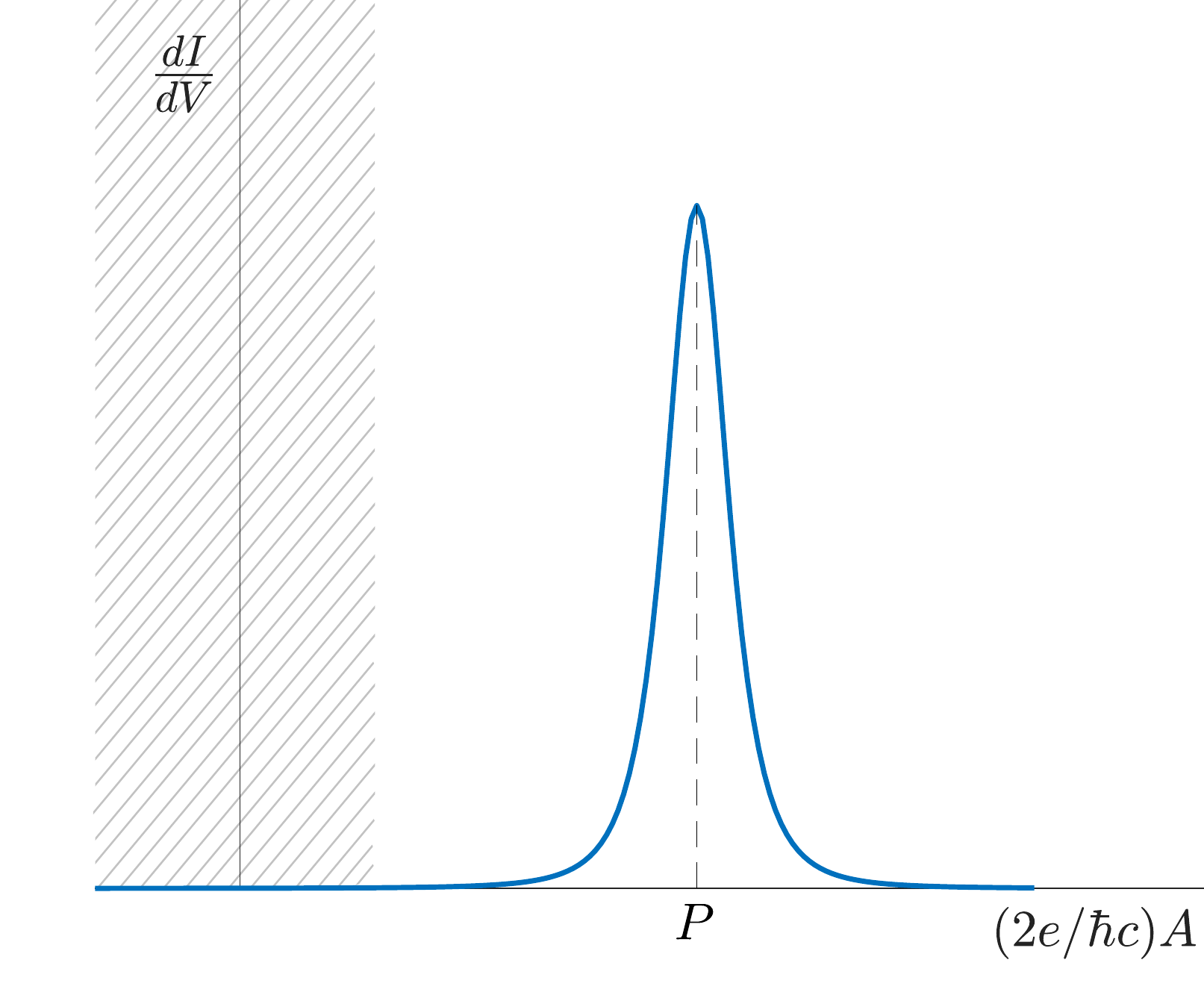}
\caption{The differential conductance measured at zero voltage is plotted against the vector potential $\mathbf{A}$ assuming a fluctuating PDW with wave-vector P in the sample. Strictly speaking the $x$ axis should be the gauge invariant gradient given by Eq. \ref{Eq: qtilde}. Here we assume there is no phase slip and take the London gauge. The shaded region denotes the $\mathbf{A}$ field that is accessible without destroying the probe superconductor. If the probe electrode has uniform pairing, a large $\mathbf{A}$ is required which is outside the accessible region. Nevertheless, even in this case there is a broad peak in $dI/dV$ as a function of voltage which may be observable for small $\mathbf{A}$. For this plot $\xi^{-1} = P/8$, ie the correlation length is about 1.27 times the PDW period.}
\label{Fig: dIdV1}
\end{center}
\end{figure}

\begin{figure}[htb]
\begin{center}
\includegraphics[width=3in]{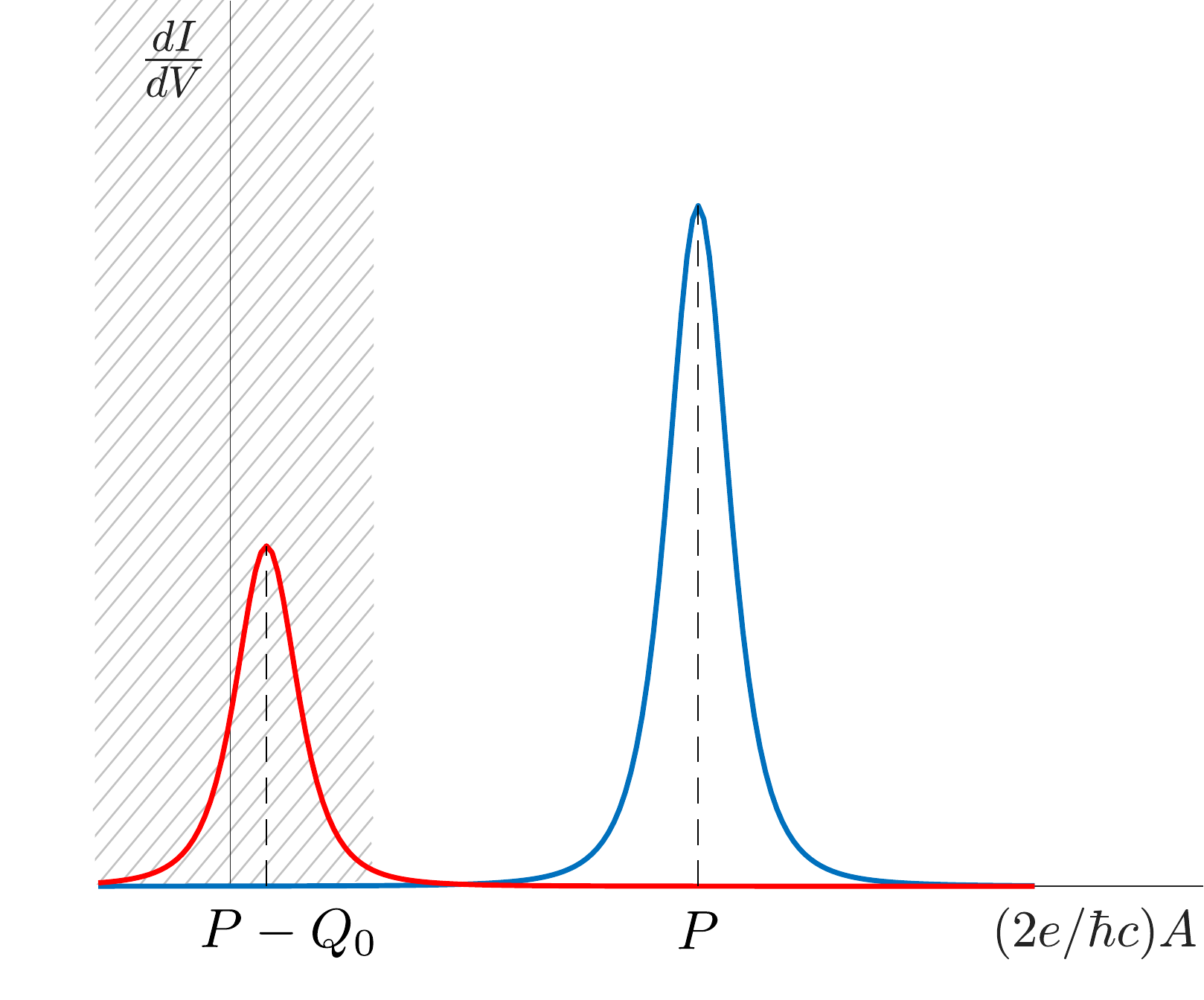}
\caption{Same as Fig 2 except that a charge order is present in the probe superconductor at wave-vector $Q_0$ which is parallel to P. This gives rise to an additional peak (in red) which partially overlaps the accessible region.}
\label{Fig: dIdV2}
\end{center}
\end{figure}

\section{General formulation.}
\noindent
We consider a tunnel barrier separating a probe superconductor and the sample we wish to study and measure the tunnel conductance G. The probe superconductor is well below its transition temperature and its order parameter is assumed to take the form 
\be
\D_0 = |\D_0|e^{-i\mathbf{Q_0}\cdot\mathbf{r}+i \phi_0(\mathbf{r})}
\label{Eq: deltaQ}
\ee
where we have assumed for generality that the probe superconductor is an ordered PDW with wave vector $\bf{Q_0}$. For conventional superconductors, we simply set $\bf{Q_0}$=0. The sample is described by a fluctuating PDW characterized by the order parameter given in Eq. \ref{Eq: deltaP}.

We assume the amplitude is fixed and the PDW is undergoing phase fluctuations. It has been shown by Scalapino \cite{scalapino} that the correlator $C(r,t)$ = $<\D_P(\mathbf{r}
,t)\D^*_P(0,0)>$ can be probed by the measuring the tunnel conductance as a function of voltage and an applied magnetic field parallel to the junction. The idea is to compute the linear response to the probe field of the ordered electrode. The voltage provides the frequency and the parallel magnetic field produces a momentum parallel to the plane for the tunneling pair , thus giving the imaginary part of the retarded pair susceptibility $\chi_R(\mathbf{q},\w)$. By the fluctuation dissipation theorem, this is related to the pair correlator. Here we extend the formulation so that it can easily describe other ways to  provide the momentum such as that shown in fig 1.  The linear response expression for the tunneling current $I$ can be written as follows
\bea
I(V)\propto &|\Delta_0|^2&\int dt \int \frac{d\mathbf{r}d\mathbf{r'}}{vol.} e^{i\omega t}e^{i\mathbf{Q_0}\cdot (\mathbf{r} - \mathbf{r'})} e^{i(\phi_0(\mathbf{r}) - \phi_0(\mathbf{r'}))}\nonumber\\
\cdot&\text{Im}& \< [ \hat{\Delta}(\mathbf{r},t), \hat{\Delta}^\dagger(\mathbf{r'},0) ] \> \theta (t)
\label{Eq: linear response1}
\eea
where $\omega=2eV/\hbar$.
By adding and subtracting in the exponent a line integral $(2e/\hbar c)\int_{\mathbf{r}}^{\mathbf{r'}}\mathbf{A}\cdot d\mathbf{l}$ along a straight line path between $\mathbf{r}$ and $\mathbf{r'}$, we rewrite this in terms of the gauge invariant response function $\chi_R(\mathbf{r-r'},t)$ 
\bea
I(V)\propto &|\Delta_0|^2&\int dt \int \frac{d\mathbf{r}d\mathbf{r'}}{vol.} e^{i\omega t} e^{i\mathbf{Q_0}\cdot (\mathbf{r} - \mathbf{r'})} e^{-i\int_{\mathbf{r}}^{\mathbf{r'}}\mathbf{\nabla}\tilde{\phi}_0(\mathbf{l})\cdot d\mathbf{l}}\nonumber\\
\cdot &\text{Im}& \chi_R(\mathbf{r} - \mathbf{r'}, t)
\label{Eq: linear response2}
\eea
where 
\bea
\chi_R(\mathbf{r} - \mathbf{r'},t) = \< [ \hat{\Delta}(\mathbf{r},t), \hat{\Delta}^\dagger(\mathbf{r'},0) ] \> \theta (t) e^{i\frac{2e}{\hbar c}\int_{\mathbf{r}}^{\mathbf{r'}}\mathbf{A}\cdot d\mathbf{l}}.
\label{Eq: chir}
\eea
It is clear that a momentum is provided in the tunneling process  by the gauge invariant phase gradient 
\be 
\tilde{\mathbf{q}}=\nabla {\tilde{\phi_0}}= \nabla \phi_0+(2e/\hbar c)\mathbf{A}
\label{Eq: qtilde}
\ee
 which is proportional to the supercurrent in the probe along the interface. From now on we shall consider the case of $\mathbf{\tilde{q}}$ being constant.  We find
\bea
I(V) \propto |\Delta_0|^2 \text{Im} \chi_R(\mathbf{q=\tilde{q}-Q_0}, \w=2eV/\hbar),
\label{Eq: IV}
\eea
where $\chi_R(\mathbf{q},\w)= \int d\mathbf{r}dt e^{-i\mathbf{q}\cdot\mathbf{r}} e^{i\omega t}\chi_R(\mathbf{r},t)$ is the Fourier transform of $\chi_R$.
 
To simplify further discussions, we shall make a simple ansatz for the response function based on the time dependent Ginzburg-Landau free energy density
$(-i\omega/\gamma_{00} + \epsilon + \xi_0^2 q^2)|\Delta(q,\omega)|^2$ where $\epsilon=(T-T_c)/T_c$. In the case of PDW fluctuations given by $\D_P$ , $q^2$ is replaced by $|\mathbf{q}+\mathbf{P}|^2$. This gives rise to
\begin{equation}
\chi_R(\mathbf{q}, \w) = [\epsilon (-i\omega/ \gamma_0 + 1 +  \xi^2 |\mathbf{q}+\mathbf{P}|^2)]^{-1}
\label{Eq: chiq}
\end{equation}
\noindent where $\gamma_0 = \epsilon \gamma_{00}$ , $\xi^2 =  \xi_0^2 /\epsilon$ are the actual inverse lifetime and correlation length of the pair fluctuations.   We shall take these as temperature dependent parameters from this point on. Since we are considering dominant phase fluctuations, we will set $\e$ as a temperature independent constant. Obviously this time dependent Ginzburg-Landau ansatz may be too simplistic to  describe the problem at hand, and one can easily use a different ansatz and follow the same principles discussed below.

Taking the imaginary part of $\chi_R$ in
 Eq.\ref{Eq: chiq} , we find the current
\begin{equation}
 I(V)=A' \w/[\gamma_B^2(1+(\w/\gamma_B)^2)]
 \label{Eq: IVgamma}
\end{equation}
where 
\be
 \gamma_B= \gamma_0 (1+\xi^2 |\mathbf{\tilde{q}}-(\mathbf{Q_0}-\mathbf{P})|^2).
 \label{Eq: gamma}
\ee
  The tunneling conductance is given by
\be
dI/dV=\frac{A\gamma_0}{\gamma_B^2}\frac{1-(\w/\gamma_B)^2}{[1+(\w/\gamma_B)^2]^2}
\label{Eq: dIdV}
\ee
where $A$ is a constant proportional to $|\D_0|^2$. In the limit of zero $\omega$, the dependence on $\mathbf{\tilde{q}}$ is in the form of the square of a Lorentzian.

\section{Application to several scenarios.}
\noindent
Now we can apply Eq.\ref{Eq: dIdV} to various set-ups to measure the pair correlator. We note that for a fixed $\mathbf{\tilde{q}}$ the voltage dependence of $dI/dV$ gives information on the lifetime of the PDW fluctuation. The lineshape is the derivative of a Lorentzian and changes sign at $\w = \gamma_B$. In the following we focus on the possibility of obtaining spatial information on the period and the decay length of the PDW by varying $\mathbf{\tilde{q}}$ and consider $dI/dV$ at zero voltage.  We begin by recalling the original Scalapino proposal where $\mathbf{\tilde{q}}$ is produced by a magnetic field parallel to the junction. As noted earlier, for a type 2 superconductor, vortex penetration of the probe electrode %
limits the $\mathbf{\tilde{q}}$ that is accessible. The set-up shown in fig 1 offers an advantage. Suppose the solenoid is turned on below the transition temperature of the probe SC, and that the sample is large enough that there is no phase slip. Let us choose  a gauge where $\bf{A}$ is along the perimeter of the hole and its magnitude
is given by $\Phi /(2\pi r)$ where   $\Phi$ is the flux through the solenoid and r is the radial distance. Since there is no phase slip,the phase winding of $\phi_0$ does not change and $\tilde{\mathbf{q}}=(2e/\hbar c)\mathbf{A}$. The SC is in a metastable state and carries a supercurrent around the hole given by the London equation which decays on a length scale given by the penetration depth. (For thin films the magnetic field generated by the screening current decays inside the hole and the total flux and its effect on $\bf{A}$ near the inner edge of the sample is small. (see reference \onlinecite{kapon2017, kerenprivate} ) There is in principle no limit to the vector potential $\bf{A}$ that can be created by the solenoid.  The only fundamental limitation is that the supercurrent may exceed the critical current and drives the probe superconductor normal. In particular,  for a d wave superconductor the supercurrent will shift the energy of the  quasi-particles by $(2e/c)\mathbf{A}\cdot \mathbf{v_F} $ where $\mathbf{v_F}$ is the Fermi velocity. Near the nodes the quasi-particles are occupied and reduce the superfluid density. This problem was treated in the paper by Yip and Sauls \cite{yip1992} and we quote their results for the reduction of the supercurrent $j_s$ in the case when $\mathbf{A}$ is directed along the direction of a node.
\be
\mathbf{j}_s= -\rho_s \mathbf{A }(1-\frac{(2e/c)|\mathbf{A}|}{2\Delta_0/v_F}).
\label{Eq: Yip}
\ee
For $\mathbf{A}$ directed along the maximum gap, there is a factor $1/\sqrt{2}$ multiplying the second term. We can think of the second term in Eq. \ref{Eq: Yip} as a reduction of the superfluid density $\rho_s$. We shall use Eq.\ref{Eq: Yip} to estimate the destruction of the probe SC. The supercurrent reaches a maximum at $\mathbf{A}$=$\mathbf{A_0}$ where $|\mathbf{A_0}|=(\hbar c/2e)\Delta_0/( \hbar v_F) = (\hbar c/2e) /\pi\xi_0$ and vanishes at 
2$|\mathbf{A_0}|$. We estimate the maximum momentum that can be accessed as $\tilde{q}=(2e/\hbar c)\mathbf{A_0}=1/(\pi \xi_0)$, ie the minimum wavelength of the PDW that can be accessed with a conventional probe SC is about $2\pi^2\xi_0$. Even with a $\xi_0$ as short as 4 lattice spacings , the minimum wavelength is about 80 lattice spacing, far too long for what we are looking for in Cuprates. Unfortunately $2\pi$ is working against us this time. The situation is illustrated in Fig 2 where we plot $dI/dV$ at zero voltage as a function of $(2e/\hbar c)A$. According to Eq.\ref{Eq: dIdV}, 
it is a Lorentzian-squared peak centered at P and the peak is far from the accessible region. It is interesting to remark that even in this case, for small $A$ , there is a peak in $dI/dV$ as a function of voltage, with a broad width of order $\gamma_B$. We suggested that this may be the broad peak seen by Koren and Lee \cite{koren2016}, but without access to the momentum dependence, it is not possible to prove or disprove this suggestion.

As seen from Eq.\ref{Eq: gamma}, this problem can be solved if we can find a probe SC which is itself a PDW with a known wave-vector $\mathbf{Q_0}$. Fortunately, we now have a candidate in the Bi-2201 family. These materials are known to have short range ordered but static charge order at wave-vector  $\mathbf{Q_0}$. \cite{comin2014, watt} By Landau theory this will induce PDW with the same wave-vector, in addition to the uniform d wave.  By choosing an appropriate optimally doped or overdoped sample as the probe SC, we can match $\mathbf{Q_0}$ close to the expected $\mathbf{P}$ of the PDW being studied and use the available $A$ range to measure the correlator in detail. This is illustrated schematically in Fig 3. Note that while the probe SC is a meta-stable state carrying a sizable  supercurrent, the pair fluctuations are in equilibrium and described by Eq.\ref{Eq: chiq}.  In the chosen gauge, we can see that the fluctuating pair will minimize their kinetic energy term by producing on average a finite $\nabla \phi$ equal to  $(2e/\hbar c)A$. This is perhaps a more physical way to understand the origin of the momentum  
$\tilde{q}$.

We remark that the $x$ axes in fig. (2) and (3) are proportional to the vector potential $A$ which can be derived from the current and the flux through the solenoid. More accurately, the x axis should be the gauge invariant phase gradient which is related to the supercurrent density. In ref \onlinecite{kapon2017} the supercurrent is directly measured via an external coil and shown to be linear in $A$ until the supercurrent collapse with the destruction of superconductivity. So in practice we can use the external coil to check that we are in the no phase slip regime.

It is also important to note that even if the sample of interest is below its transition temperature for long range order of the uniform SC, if the PDW continues to be fluctuating, it remains in equilibrium and its contribution to $dI/dV$ does not change. So it is possible to zero field cool to a temperature where the sample has conventional uniform long range ordered superconductivity. The usual Josephson current can be suppressed by a small in plane magnetic field and the PDW spectrum can be probed.

Next we discuss the case when the PDW in the sample also has long range order, as is possibly the case in LBCO below 30K. If the solenoid is turned on below its ordering temperature, the sample also carries a metastable supercurrent and its susceptibility is shifted by $\tilde{q}$ from that given in Eq \ref{Eq: chiq}. As a result $\tilde{q}$ is canceled in Eq.\ref{Eq: qtilde} and $dI/dV$ is independent of the solenoid field. Physically when both the probe and the sample are in a metastable current carrying state, the effect of the vector potential $\mathbf{A}$ is canceled. There is still a Josephson current when $Q_0$ happens to match $P$, but it may be difficult to distinguish it from the conventional Josephson current due to the uniform SC in the probe and the sample. However, the  $\mathbf{A}$ dependence can be restored with the following protocol. Turn on the solenoid between the two ordering temperatures and cool to below the sample 
$T_c$ to make the $dI/dV$ measurement. Then warm up to a temperature between the two $T_c$'s, change the solenoid current and cool down again below the sample $T_c$ to measure $dI/dV$. In this way the sample is always in a field cooled state and carry no supercurrent while the probe remains in the zero field cooled metastable state. Hence, the dependence on $\tilde{q}$ in \ref{Eq: chiq} is restored. The different behavior between the two protocols described above is a signature of the ordered PDW. 

Finally I mention a number of experimental complications that need to be accounted for in the data analysis. In Bi-2201 there is a distribution of charge order wave-vector, hence a distribution in $Q_0$. \cite{watt} Averaging over this distribution will broaden the $dI/dV$ curves and the curves shown in figs (2,3). This broadening actually works to our advantage in the case of ordered PDW as in LBCO samples, because the Josephson current will appear without too much fine-tuning the matching between $P$ and $Q_0$. As a first step, one may attempt to see this Josephson current without using the solenoid.  Secondly the vector potential $\mathbf{A}$ rotates in direction as one moves around the perimeter of the hole and is not always lined up with $\mathbf{P}$. The vector nature of the $\tilde{q}$ in Eq. \ref{Eq: gamma} needs to be taken care of and an appropriate average made. Along the same vein, we have to average over the $\mathbf{-P}$ and $\mathbf{-Q_0}$ contributions and include the fact that the charge order and therefore the PDW order in the probe SC  runs along both x and y directions. Also we have to be mindful of the  possibility that the fluctuating PDW in the sample is bi-directional. Similarly the  radial dependence of $\mathbf{A}$ needs to be accounted for; in practice  the tunneling current of interest comes only from the region within the London penetration depth of the inner edge of the sample.  All these complications can be accounted for by appropriate averaging of the basic equation \ref{Eq: dIdV}. Finally we have to make sure there is no phase slip in the probe SC. This may be aided by the introduction of vortex pins.

While we have emphasized in this paper the use of the solenoid to provide information on the momentum dependence of the fluctuation spectrum, we would like to remind the readers that valuable information is also gained from the voltage dependence, which probe the time dependence of the PDW fluctuations. Since the momentum match can be provided by an appropriately chosen probe SC, as a first step one can do without the solenoid and simply study the voltage dependence of a tunnel junction with a suitable Bi-2201 electrode as the probe SC.

In conclusion, the geometry shown in Fig 1 combined with the use of an appropriate Bi-2201 as probe tunneling electrode allows us to get quite detailed information on the spatial and temporal information of a fluctuating or ordered PDW. This experiment should allow us to directly confirm the existence of an ordered stripe PDW in LBCO. In the YBCO and the Bi-2201 and Bi-2212 family, it will be most interesting to study the metallic phase once the uniform d wave order is suppressed by a large supercurrent. It is known that large diamagnetic signals remain for field larger than the resistive $H_{c2}$,  \cite{yuPNAS126672016magnetic} and it will be good to know whether this is coming from fluctuating uniform d-wave or from fluctuating PDW. \cite{lee2014amperean} The method proposed in this paper should allow us to decide between these possibilities.

{\em Acknowledgments:}
I thank Assa Auerbach for bring the work of Kapon et al.  \cite{kapon2017} to my attention. I thank Amit Keren for sharing information from his unpublished data. I thank Gad Koren and Zhehao Dai for helpful discussions. This work has been supported by DOE office of Basic Sciences grant number DE-FG02-03-ER46076.

\bibliographystyle{apsrev4-1_custom}
\bibliography{test}

\end{document}